\documentclass[sigconf,nonacm,10pt]{acmart}
\usepackage[english]{babel}
\usepackage{booktabs}
\usepackage{subcaption}
\usepackage{graphicx}

\acmDOI{}

\author{Shuyao Qi}
\affiliation{%
  \institution{Shanghai Jiao Tong University}
  \city{Shanghai}
  \country{China}}
\email{nastyapple@sjtu.edu.cn}

\author{Haoyuan Liu}
\affiliation{%
  \institution{Shanghai Jiao Tong University}
  \city{Shanghai}
  \country{China}}
\email{cse-lhy@sjtu.edu.cn}

\author{Shizhen Zhao}
\affiliation{%
  \institution{Shanghai Jiao Tong University}
  \city{Shanghai}
  \country{China}}
\email{shizhenzhao@sjtu.edu.cn}

\renewcommand{\shortauthors}{Shuyao Qi, Haoyuan Liu, Shizhen Zhao}

\begin{document}

\title{FEPLB: Exploiting Copy Engines for Nearly Free MoE Load Balancing in Distributed Training}

\begin{abstract}
Fine-grained, per-micro-batch load balancing is essential for efficient Mixture-of-Experts (MoE) training, yet every prior dynamic scheduling scheme pays for it with extra communication that is hard to hide---especially on modern bulk-transfer backends such as DeepEP. We make a simple but consequential observation: on the NVIDIA Hopper architecture the NVLink Copy Engine can move data between intra-node GPUs without consuming any SM cycles, effectively providing a nearly free communication channel that runs in parallel with compute kernels. FEPLB turns this idle hardware into a new parallel dimension for MoE load rebalancing. Its Two-Phase Dispatch first routes tokens across nodes via the standard EP backend, then redistributes dynamic-expert tokens and weights within the NVLink domain through the Copy Engine at nearly zero cost, while a lightweight CPU scheduler runs concurrently with static expert computation. Because FEPLB uses only Copy Engine and CPU that are orthogonal to those consumed by EP and PP, it coexists with existing parallel strategies without reconfiguration.

On GLM-5's MoE layers (128 experts, no auxiliary loss, up to 16 H100 GPUs), FEPLB reduces the token straggler by 51--70\% and the GEMM straggler by 50--68\% with no measurable EP communication overhead. Its advantage grows with the EP degree: at EP\,=\,8, it achieves 2$\times$ lower token straggler than FasterMoE.
\end{abstract}


\maketitle

\section{Introduction}
\label{sec:intro}

Mixture-of-Experts (MoE) has become the dominant architecture for scaling large language models. Models such as DeepSeek-V3~\cite{liu2024deepseek}, Kimi-K2~\cite{team2025kimi}, and GLM-5~\cite{zeng2025glm} reach trillions of parameters while activating only a small fraction per token. Under Expert Parallelism (EP), each device holds a disjoint subset of experts; the per-device computational load depends on how many tokens the learned router assigns, a quantity that varies randomly per micro-batch.

\begin{figure*}[t]
  \centering
  \includegraphics[width=\linewidth]{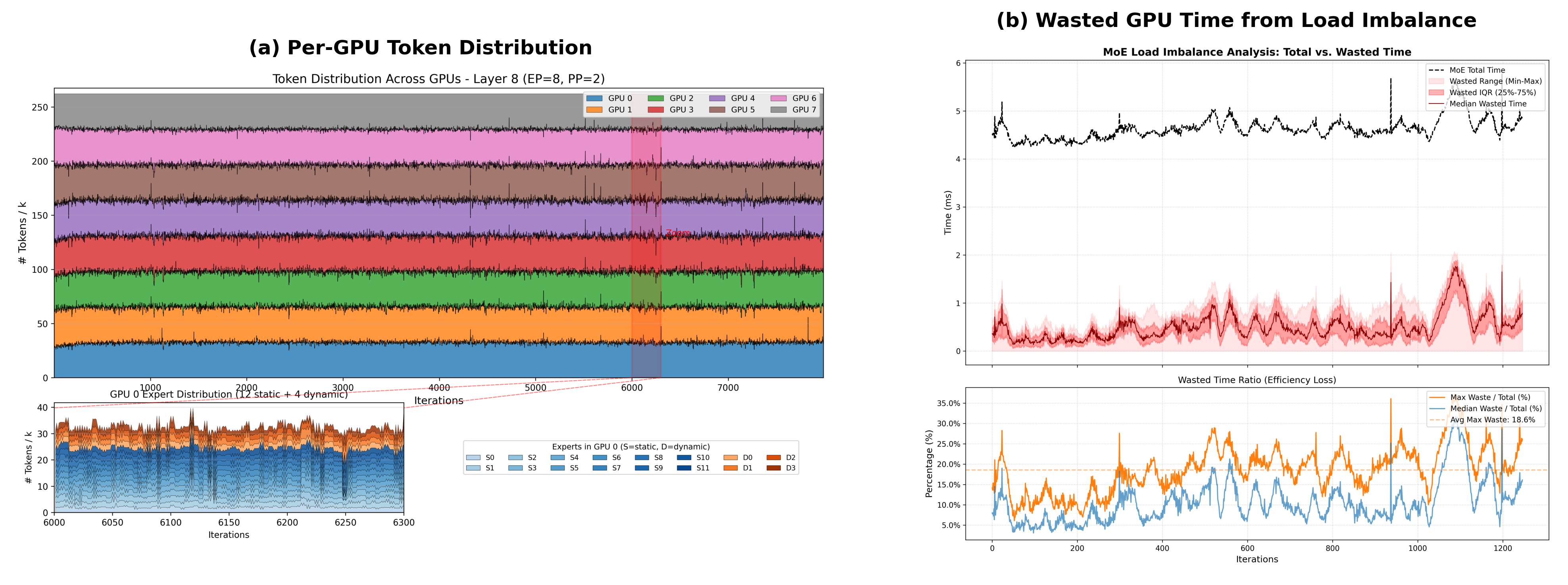}
  \caption{(a) Per-GPU token distribution (stacked) across 7{,}000 training iterations for GLM-5's MoE layer (128 experts, EP\,=\,8, without auxiliary loss). The varying band widths reveal persistent, random load imbalance across all eight GPUs. (b) Wasted GPU time analysis: the upper panel shows total MoE layer time vs.\ wasted time (the gap between the slowest and average device); the lower panel shows the efficiency loss ratio. On average, load imbalance wastes 18.6\% of GPU time per MoE layer.}
  \label{fig:expert_load}
\end{figure*}

Modern distributed training employs parallel dimensions (Data Parallelism (DP), Tensor Parallelism (TP), EP, Pipeline Parallelism (PP)), all configured offline assuming deterministic tensor shapes and communication volumes~\cite{zheng2022alpa, unger2022unity, miao2022galvatron, tensoropt}. MoE routing breaks this assumption. As Figure~\ref{fig:expert_load}(a) shows, per-device token counts fluctuate widely across micro-batches even during stable training, and the imbalance pattern is data-dependent: different training corpora and stages produce different routing distributions. Coarse-grained mitigations such as auxiliary balancing losses~\cite{fedus2022switch, lepikhin2020gshard} constrain the router's expressiveness, degrading model quality. The alternative, accepting the imbalance, leads to significant wasted compute: as Figure~\ref{fig:expert_load}(b) shows, in synchronous training every device waits for the slowest, and load imbalance wastes on average 18.6\% of GPU time per MoE layer.

Prior work on dynamic MoE scheduling attempts to address this, but fails to fully hide the overhead of rebalancing. Tutel~\cite{hwang2023tutel} and SmartMoE~\cite{zhai2023smartmoe} switch between predefined parallel configurations, but their strategies partition expert weights across GPUs, always introducing additional communication. FasterMoE~\cite{he2022fastermoe} replicates hot experts within the EP domain (shadow expert) and pipelines dispatch with computation; however, pipelining and shadow expert splits Grouped GEMM (general matrix multiply) into smaller per-expert kernels, degrading GPU utilization, and its prediction-based scheduling degrades under changing routing patterns. Triton Distributed~\cite{triton_distributed} fuses computation with communication for TP-parallel MoE, but the overlap consumes SM resources, reducing compute efficiency.

A deeper compatibility issue is common to all these approaches. In practice, specialized MoE communication libraries such as DeepEP~\cite{deepep} and FUSCO~\cite{fusco} deliver much better performance than All-to-All collectives through custom dispatch protocols. Importantly, these libraries perform bulk token transfers without support for staged delivery. This breaks the staged pipelining assumed by prior overlap-based approaches: splitting communication into stages on top of these backends adds extra volume rather than hiding latency (Section~\ref{sec:ep_comm}).


We observe that on the NVIDIA Hopper architecture, the NVLink Copy Engine combined with symmetric memory provides a path for intra-node communication that is entirely invisible to GPU computation. Unlike SM-based communication (NCCL, fused kernels) that competes for compute resources, the Copy Engine operates on a dedicated hardware data path and proceeds through NVLink without consuming any SM cycles or interfering with concurrent Grouped GEMM execution. This hardware capability has been mostly unused for MoE load balancing.

Building on this observation, we design FEPLB, a system that introduces a new parallel dimension dedicated to dynamic, per-micro-batch load scheduling. FEPLB uses Two-Phase Dispatch: Phase 1 routes tokens to the NVLink domain via EP, with static expert tokens dispatched normally and dynamic expert tokens collected for intra-node processing; Phase 2 redistributes tokens and copies expert weights within the NVLink domain via the Copy Engine at nearly zero cost. A CPU-side load balancer decides per micro-batch which experts to rebalance, running concurrently with static expert computation. Because the Copy Engine, CPU scheduling, and the time during static expert computation are all resources unused by EP and PP, FEPLB achieves orthogonality by construction: it coexists with existing parallel dimensions without reconfiguring parallelism or replacing the EP communication backend.

We evaluate FEPLB on the MoE layers of GLM-5 (128 routed experts, without auxiliary loss) across three PP/EP configurations on up to 16 NVIDIA H100 GPUs. Our contributions are:

\begin{enumerate}
  \item We propose dynamic load balancing as a new parallel dimension for MoE, orthogonal to EP and PP, and identify the design principle that enables its introduction: resource-level separation from existing parallel domains.
  \item We design Two-Phase Dispatch, which separates inter-node EP routing (Phase 1) from SM-free, NVLink-CE-based intra-node load rebalancing that redistributes both tokens and expert weights (Phase 2), achieving orthogonality by construction.
  \item We present a per-micro-batch expert weight redistribution algorithm that runs on the CPU concurrently with static expert computation, with a configurable dynamic expert count (dyn) and minimum token threshold.
  \item We demonstrate that FEPLB reduces token straggler by 51--70\% and GEMM straggler by 50--68\% with no measurable EP communication overhead. Its advantage grows with EP degree, achieving 2$\times$ lower token straggler than FasterMoE at EP\,=\,8.
\end{enumerate}

\section{Design of FEPLB}

\subsection{Design Principle: Orthogonal Dynamic Parallelism}
\label{sec:principle}

The central design principle is that FEPLB's parallel dimension must be orthogonal to EP and PP. Concretely: (1) EP communication must be unaffected, with no additional inter-node traffic or NIC interference; (2) PP scheduling must be unchanged, as FEPLB operates within a single MoE layer; (3) GPU SMs must not be consumed, so all rebalancing communication must be SM-free.

\begin{table}[t]
  \caption{Parallel dimensions in MoE training and their hardware resource usage. FEPLB introduces a new dynamic dimension that uses resources (CPU, NVLink Copy Engine) orthogonal to those used by EP and PP.}
  \label{tab:dimensions}
  \centering
  \begin{tabular}{@{}llll@{}}
    \toprule
    Dimension & Scope & Communication & Compute \\
    \midrule
    EP        & Inter-node  & RDMA / NVLink      & GPU SMs \\
    PP        & Inter-node  & RDMA / NVLink     & GPU SMs \\
    FEPLB & Intra-node & NVLink Copy Engine & CPU \\
    \bottomrule
  \end{tabular}
\end{table}

Table~\ref{tab:dimensions} shows how FEPLB achieves this via resource-level separation. EP and PP both use RDMA/NVLink and GPU SMs; FEPLB uses NVLink Copy Engine and CPU. These resource sets do not overlap, making FEPLB a true parallel dimension rather than an ad-hoc scheduling trick.

\subsection{Two-Phase Dispatch}
\label{sec:twophase}

\begin{figure}[t]
  \centering
  \includegraphics[width=\linewidth]{figures/two-phase-dispatch-for-moe-dim.png}
  \caption{Two-Phase Dispatch. Phase 1: EP dispatch routes static expert tokens to assigned devices and collects dynamic expert tokens into the NVLink domain (inter-node, $\sim$50\,GB/s). Phase 2: NVLink Copy Engine redistributes dynamic expert tokens and copies expert weights within the node (SM-free, $\sim$900\,GB/s). No cross-node balancing is performed, preserving EP's inter-node communication pattern.}
  \label{fig:twophase}
\end{figure}

Two-Phase Dispatch decomposes MoE token processing into two stages (Figure~\ref{fig:twophase}).

In Phase 1, static expert tokens are dispatched to their assigned devices via the standard EP backend (e.g., DeepEP), following the unmodified EP path. Dynamic expert tokens are collected from the entire EP domain into the corresponding NVLink domain (i.e., the local node) and permuted there, preparing them for intra-node redistribution.

After Phase 1, each device within the NVLink domain knows the exact per-expert token counts for dynamic experts, which may be highly unbalanced across devices.

In Phase 2, the NVLink Copy Engine redistributes dynamic expert tokens and copies expert weights from overloaded to underloaded devices within the NVLink domain, consuming zero SM cycles. The Copy Engine operates at 900\,GB/s bidirectional on H100 NVLink 4.0, on a hardware path separate from both the inter-node NICs used by EP and the SMs used for Grouped GEMM. FEPLB rebalances only within the NVLink domain, leaving EP's inter-node communication untouched. Because each token is still processed by the same expert with identical weights, weight redistribution preserves exact MoE semantics.

This scope is limited by the current NVLink topology, not by design. On NVIDIA SuperPod architectures (e.g., GB200 NVL72) with all-to-all NVLink across 72 GPUs, Phase 2 can rebalance the entire EP group without cross-node communication.

\subsection{System Architecture and Overlap Strategy}
\label{sec:system}

\begin{figure}[t]
  \centering
  \includegraphics[width=\linewidth]{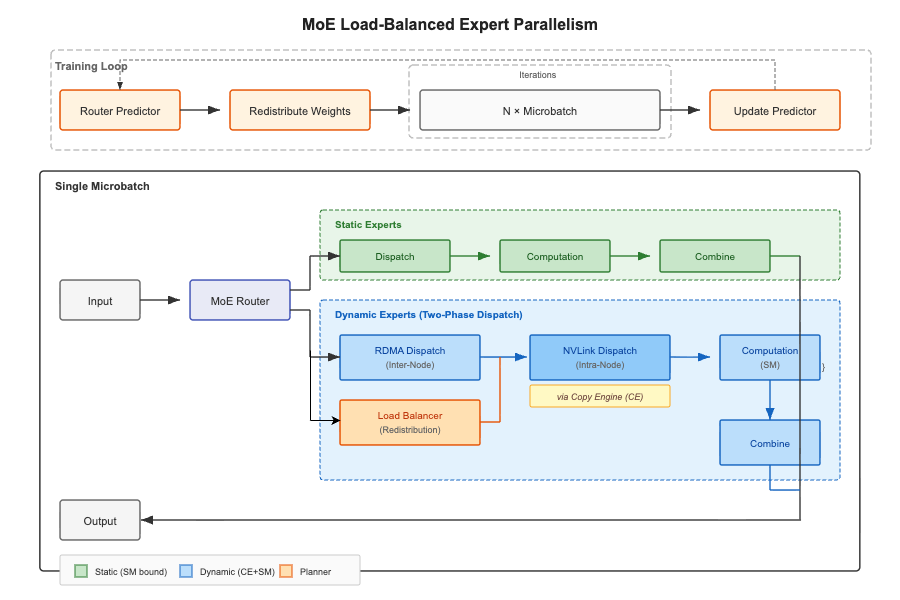}
  \caption{FEPLB system architecture. The training loop (top) periodically optimizes expert placement via a Router Predictor. Within each micro-batch (bottom), static experts follow the standard EP path while dynamic experts use Two-Phase Dispatch with SM-free NVLink weight redistribution.}
  \label{fig:system}
\end{figure}

FEPLB operates at two timescales (Figure~\ref{fig:system}).

At the macro level, a Router Predictor periodically optimizes expert-to-device assignment based on historical routing statistics, executed at checkpoint time to spread out migration cost.

At the micro level, each micro-batch uses Two-Phase Dispatch for fine-grained rebalancing. Experts on each device are partitioned into static and dynamic categories, controlled by parameter $\text{dyn}$. With 128 experts and EP\,=\,8, each device hosts 16; setting dyn\,=\,4 makes 4 dynamic (eligible for copying) and 12 static (always local).

The per-micro-batch timeline exploits this partition. First, the router produces routing decisions on all devices. Phase 1 then dispatches static expert tokens to their assigned devices via the standard EP backend while collecting dynamic expert tokens from the EP domain into the NVLink domain and permuting them. The GPU then begins computing static experts on SMs; concurrently, the CPU load balancer analyzes the actual token distribution and decides which dynamic experts' weights to copy, and the NVLink Copy Engine transfers those weights along with the corresponding permuted tokens to target devices, all SM-free. Once static expert computation and weight copy both complete, the GPU computes dynamic experts with rebalanced load. Finally, the Combine phase returns results to source devices.

Static experts serve a dual purpose: they contribute to the model's output, and their computation provides a time window during which the load balancer and weight copy finish without blocking the critical path. The CPU load balancer runs on a dedicated thread, triggered by router completion, and issues weight copy commands to the NVLink Copy Engine on dedicated CUDA streams without SM involvement.

The load balancer runs a greedy algorithm: it repeatedly selects the busiest dynamic expert on the most overloaded device and copies that expert's weights (along with its tokens) to the most underloaded device. A minimum token threshold $\tau$ prevents copying experts with too few tokens. Each copy migrates an entire expert rather than splitting its token batch: Grouped GEMM performance is highly sensitive to per-expert batch size under the roofline model, and splitting tokens would produce smaller matrix multiplications in the memory-bound regime. Migrating whole experts also makes the algorithm deterministic: given the same routing decisions, every device independently derives the same weight copy plan without coordination. The algorithm completes in approximately 50\,$\mu$s on a single CPU core, well within the static expert computation window.

The memory overhead is modest: FEPLB allocates $\text{max\_num\_dyn} \times W_{\text{expert}}$ per device for copied weights. For GLM-5, each expert is 72\,MiB; with max\_num\_dyn\,=\,8, the buffer is 576\,MiB ($<$0.7\% of 80\,GB HBM3), reused across all MoE layers.

\section{Evaluation}

We evaluate FEPLB on per-layer execution time, EP communication overhead, load balance quality across methods and EP degrees, and sensitivity to the dynamic expert count.

\subsection{Experimental Setup}
\label{sec:setup}

\paragraph{Hardware.}
Experiments run on NVIDIA H100 SXM5 GPUs (80\,GB HBM3) connected via NVLink 4.0 (900\,GB/s bidirectional per GPU) within each node and 400\,Gbps InfiniBand between nodes.

\paragraph{Software framework.}
Our baseline is the Megatron-LM MoE framework with DeepEP for dispatch, NVIDIA Transformer Engine for mixed-precision training, and multi-stream cuBLAS-based Grouped GEMM. FEPLB is implemented on top; all configurations share the same communication and computation kernels.

\paragraph{Model configuration.}
We use a reduced-layer variant of GLM-5~\cite{zeng2025glm} (18 layers instead of the original 78), preserving the original MoE layer architecture (128 routed experts, top-$k$ routing, no auxiliary loss). Since FEPLB operates independently within each MoE layer, reducing the layer count does not affect per-layer evaluation.

We evaluate three PP/EP configurations covering different parallelism trade-offs:
PP\,=\,4, EP\,=\,2 (8 GPUs);
PP\,=\,4, EP\,=\,4 (16 GPUs);
PP\,=\,2, EP\,=\,8 (16 GPUs).
Each configuration assigns $128 / \text{EP}$ experts per device (64, 32, or 16 respectively).

\paragraph{Baselines.}
We compare five configurations:
(1) Before LB: standard EP with no load balancing;
(2) FasterMoE~\cite{he2022fastermoe}: shadow-expert replication with pipe\,=\,1 (no overlap) and pipe\,=\,2 (pipelined dispatch). For fair comparison, we re-implement FasterMoE with SM-free NVLink CE transfers and DeepEP dispatch. Unless noted, results refer to pipe\,=\,1;
(3) Triton Distributed~\cite{triton_distributed}: TP-parallel MoE with fused computation-communication kernels;
(4) Tutel~\cite{hwang2023tutel}: adaptive switching between EP and DP modes;
(5) FEPLB: our system with Two-Phase Dispatch.

\paragraph{Metrics.}
We report two straggler metrics that directly quantify load imbalance.
The token straggler is $\max_d T_d - \bar{T}$, where $T_d$ is the per-GPU token count and $\bar{T}$ is the mean. It measures the excess token count on the most loaded device.
The GEMM straggler is $\max_d G_d - \bar{G}$, where $G_d$ is the per-GPU Grouped GEMM execution time. It measures the wall-clock computation time wasted waiting for the slowest device.
Both metrics are averaged over the full training run.

\subsection{Per-Layer Execution Time}
\label{sec:per_layer}

\begin{table}[t]
  \caption{Per-layer MoE execution time (forward / backward, ms) across PP/EP configurations under five methods.}
  \label{tab:per_layer}
  \centering
  \resizebox{\columnwidth}{!}{
  \begin{tabular}{@{}lccccc@{}}
    \toprule
    PP/EP & Before LB & FasterMoE & Triton Dist. & Tutel & FEPLB \\
    \midrule
    4/2 & 8.2/14.9 & \textbf{7.9}/\textbf{14.0} & 13.1/22.8 & 8.0/17.1 & \textbf{7.9}/14.4 \\
    4/4 & 7.3/13.2 & 6.9/12.2 & 15.3/24.0 & 7.2/15.2 & \textbf{6.8}/\textbf{12.1} \\
    2/8 & 6.9/12.5 & 6.3/11.1 & 22.8/30.0 & 6.8/14.5 & \textbf{6.0}/\textbf{10.6} \\
    \bottomrule
  \end{tabular}
  }
\end{table}

Table~\ref{tab:per_layer} reports per-MoE-layer execution time. Triton Distributed is 1.6--3.3$\times$ slower than the baseline on the forward pass: its fused computation-communication kernels consume SM resources for communication, and the overhead grows as more GPUs participate in the collective. Tutel matches or slightly improves forward time but adds 15--16\% backward overhead due to additional communication from weight partitioning across GPUs.

FEPLB consistently matches or outperforms all baselines. At PP\,=\,2, EP\,=\,8, forward time drops from 6.9\,ms to 6.0\,ms ($-$13\%) and backward from 12.5\,ms to 10.6\,ms ($-$15\%). At PP\,=\,4, EP\,=\,2, FEPLB's forward matches FasterMoE (7.9\,ms) but backward is slightly higher (14.4 vs.\ 14.0\,ms): with EP\,=\,2, each device hosts 64 experts, and since FEPLB migrates entire experts to preserve Grouped GEMM efficiency, rebalancing granularity is limited at low EP.

\subsection{Orthogonality Verification: EP Communication Overhead}
\label{sec:ep_comm}

\begin{figure}[t]
  \centering
  \includegraphics[width=\linewidth]{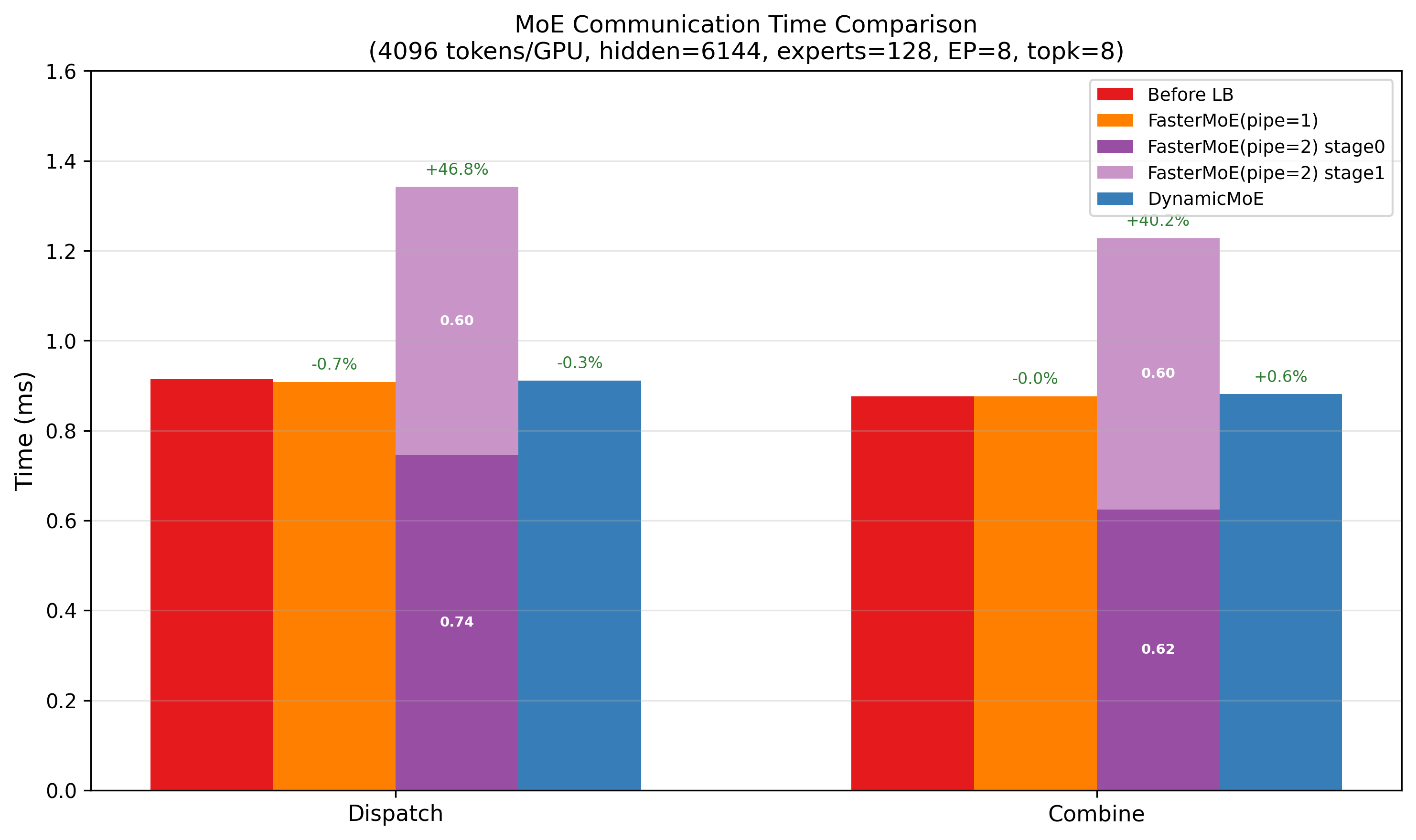}
  \caption{EP communication time (Dispatch and Combine) for Before LB, FasterMoE (pipe\,=\,1 and pipe\,=\,2), and FEPLB. FasterMoE with pipe\,=\,2 adds up to 46.8\% dispatch overhead due to additional communication volume on top of DeepEP. FEPLB introduces no measurable overhead ($<$1\%).}
  \label{fig:ep_comm}
\end{figure}

Figure~\ref{fig:ep_comm} tests whether each approach alters EP communication performance (measured at EP\,=\,8). FasterMoE with pipe\,=\,1 shows negligible overhead, matching FEPLB. However, pipe\,=\,2 incurs 46.8\% additional dispatch time and 40.2\% additional combine time, validating the observation from Section~\ref{sec:intro}: pipelining requires splitting communication into stages, which adds volume on top of bulk-transfer backends like DeepEP. FEPLB avoids this: Phase 2 operates after Phase 1 on a separate hardware path, with overhead below 1\%.

\subsection{Load Balance Quality}
\label{sec:balance}

\begin{figure}[t]
  \centering
  \includegraphics[width=\linewidth]{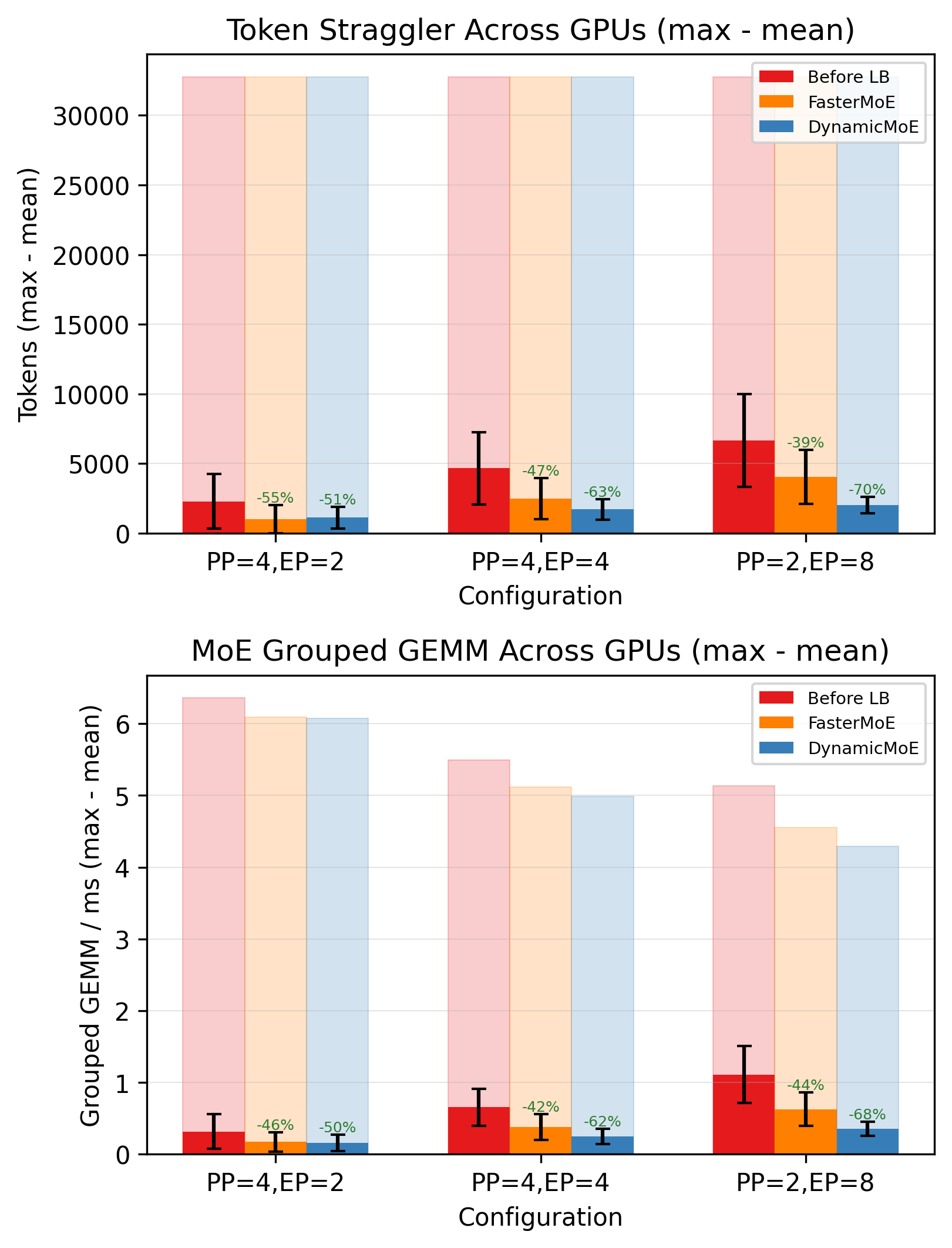}
  \caption{Token straggler (top) and GEMM straggler (bottom) across three PP/EP configurations for Before LB, FasterMoE, and FEPLB. FEPLB achieves 51--70\% token straggler reduction and 50--68\% GEMM straggler reduction, increasingly outperforming FasterMoE as EP grows.}
  \label{fig:shadow}
\end{figure}

Figure~\ref{fig:shadow} compares load balance quality across all PP/EP configurations.

\begin{table}[t]
  \caption{Token straggler (max $-$ mean) across configurations. Percentages show reduction relative to Before LB.}
  \label{tab:token_straggler}
  \centering
  \begin{tabular}{@{}lccc@{}}
    \toprule
    PP/EP & Before LB & FasterMoE & FEPLB \\
    \midrule
    4/2 & 2{,}278 & 1{,}014 ($-$55\%) & 1{,}107 ($-$51\%) \\
    4/4 & 4{,}649 & 2{,}471 ($-$47\%) & 1{,}697 ($-$63\%) \\
    2/8 & 6{,}666 & 4{,}036 ($-$39\%) & 2{,}021 ($-$70\%) \\
    \bottomrule
  \end{tabular}
\end{table}

\begin{table}[t]
  \caption{GEMM straggler in ms (max $-$ mean) across configurations. Percentages show reduction relative to Before LB.}
  \label{tab:gemm_straggler}
  \centering
  \begin{tabular}{@{}lccc@{}}
    \toprule
    PP/EP & Before LB & FasterMoE & FEPLB \\
    \midrule
    4/2 & 0.316 & 0.170 ($-$46\%) & 0.157 ($-$50\%) \\
    4/4 & 0.652 & 0.380 ($-$42\%) & 0.247 ($-$62\%) \\
    2/8 & 1.110 & 0.625 ($-$44\%) & 0.352 ($-$68\%) \\
    \bottomrule
  \end{tabular}
\end{table}

FEPLB and FasterMoE exhibit opposite scaling trends. At EP\,=\,2, FasterMoE achieves slightly better token straggler reduction (55\% vs.\ 51\%) thanks to relatively stable routing at low EP, though FEPLB already edges ahead on GEMM straggler (50\% vs.\ 46\%). As EP increases, FasterMoE's prediction accuracy degrades under sparser, less predictable routing distributions: its token straggler reduction drops from 55\% to 39\%. FEPLB's reactive approach improves in the opposite direction, from 51\% to 70\%. At EP\,=\,8, FEPLB achieves 2$\times$ lower token straggler (2{,}021 vs.\ 4{,}036) and 1.8$\times$ lower GEMM straggler (0.352 vs.\ 0.625\,ms) than FasterMoE.

\subsection{Sensitivity to Dynamic Expert Count}
\label{sec:dyn}

\begin{figure}[t]
  \centering
  \includegraphics[width=\linewidth]{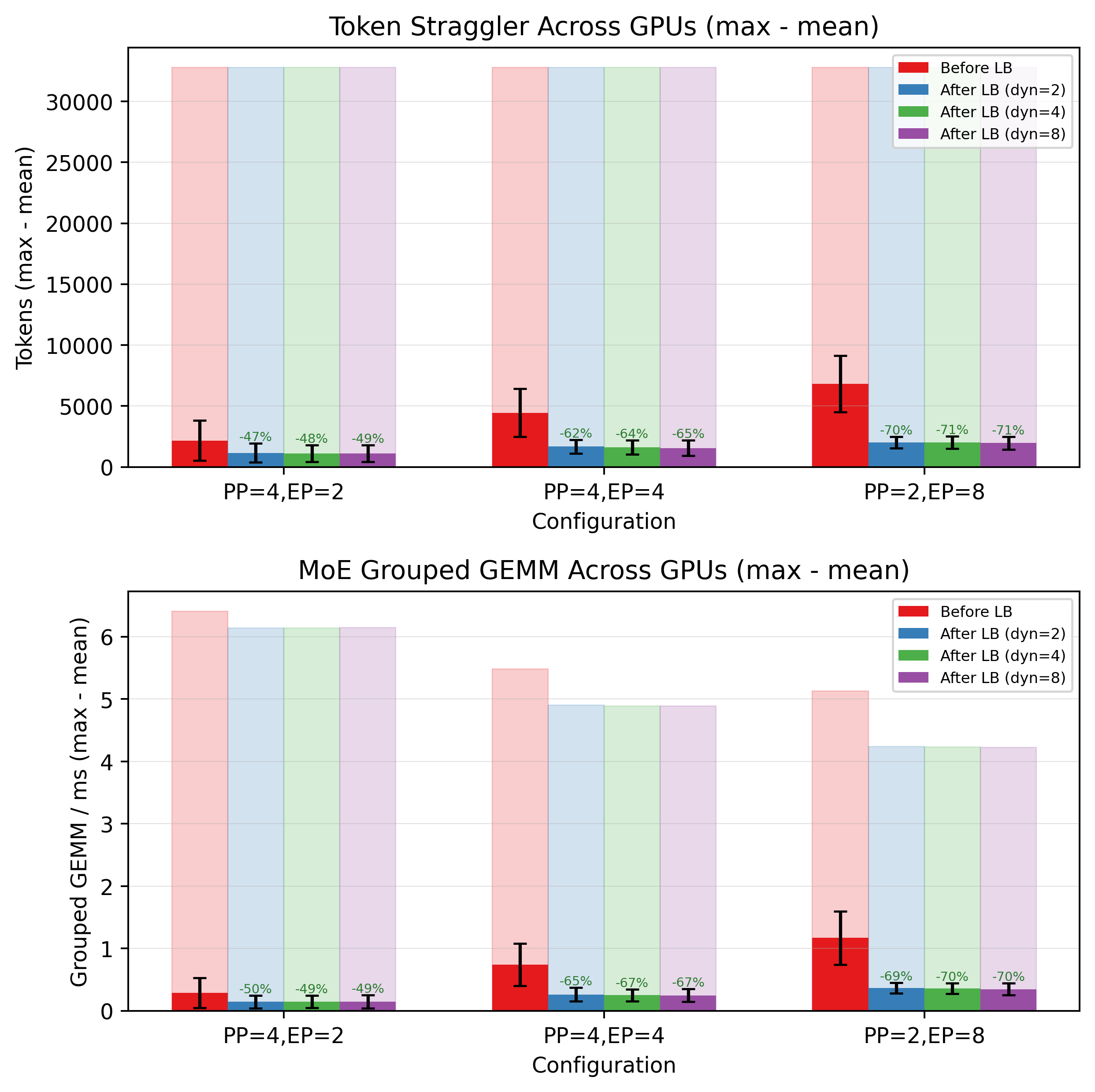}
  \caption{Token straggler as a function of the dynamic expert count dyn\,=\,2, 4, 8 across three PP/EP configurations. Higher dyn provides more rebalancing flexibility, with diminishing returns beyond dyn\,=\,4.}
  \label{fig:dyn}
\end{figure}

The parameter dyn controls how many experts per device are eligible for dynamic copying. Figure~\ref{fig:dyn} evaluates dyn\,=\,2, 4, 8.

Even dyn\,=\,2 achieves substantial token straggler reduction, because the imbalance is usually caused by a few busy experts. Increasing dyn yields diminishing returns: dyn\,=\,2 to 4 adds 1--3 percentage points, and 4 to 8 another 1--3 points. dyn\,=\,4 is a practical default.

\section{Related Work}

\paragraph{MoE systems and parallel strategies.}
GShard~\cite{lepikhin2020gshard}, Switch Transformer~\cite{fedus2022switch}, DeepSeek-V3~\cite{liu2024deepseek}, and GLM-5~\cite{zeng2025glm} scaled MoE to trillions of parameters, while Megatron-LM~\cite{shoeybi2019megatron} and DeepSpeed-MoE~\cite{rajbhandari2022deepspeed} built distributed training infrastructure. Auto-parallelization frameworks~\cite{zheng2022alpa, unger2022unity, miao2022galvatron, tensoropt, shi2023tap, she2025automatic, li2024automatically} solve for parallel strategies offline under deterministic assumptions violated by MoE routing. FEPLB complements these systems by dynamically reconfiguring within each MoE layer.

\paragraph{Dynamic scheduling and communication.}
Tutel~\cite{hwang2023tutel} switches between EP and DP modes but partitions weights, adding communication. SmartMoE~\cite{zhai2023smartmoe} selects from a pre-computed strategy menu. FasterMoE~\cite{he2022fastermoe} replicates hot experts; even re-implemented with SM-free NVLink CE and DeepEP, its predictive approach degrades as EP increases (Section~\ref{sec:balance}). Triton Distributed~\cite{triton_distributed} fuses TP-parallel MoE with communication but reduces available SMs. Specialized libraries such as DeepEP~\cite{deepep} and FUSCO~\cite{fusco} perform bulk transfers without staged delivery, breaking the pipelining assumed by prior overlap approaches. FEPLB is compatible with these backends and differs by reconfiguring every micro-batch through resource-level separation.

\section{Conclusion}

We present FEPLB, a system that introduces dynamic load balancing as a new parallel dimension orthogonal to EP and PP via resource-level separation: NVLink Copy Engine for intra-node redistribution, CPU for scheduling, separate from the RDMA NICs and GPU SMs used by existing dimensions. Two-Phase Dispatch separates inter-node EP routing from SM-free intra-node rebalancing, preserving exact MoE semantics and supporting fully auxiliary-loss-free training.

On GLM-5's MoE layer (128 experts, no auxiliary loss, up to 16 H100 GPUs), FEPLB reduces token straggler by 51--70\% and GEMM straggler by 50--68\% with no measurable EP overhead, achieving 2$\times$ lower token straggler than FasterMoE at EP\,=\,8. Current limitations include whole-expert migration without token-level splitting, which limits granularity at low EP. This work shows that the dynamic overheads of conditional computation can be eliminated by using hardware resources that current frameworks leave idle.

\bibliographystyle{ACM-Reference-Format}

\bibliography{reference}

\end{document}